\documentstyle[aps]{revtex}
\begin{document}
\bibliographystyle{prsty}

\title{\Large\bf General Multimode Squeezed States}

\author{{Gan Qin$^{1,2}$\thanks{E-Mail: gqin@ustc.edu.cn},
Ke-lin Wang$^{2}$, and Tong-zhong Li$^{1,3}$}\\
$^1${\small\it CCAST (World Lab.), P. O. Box 8730, Beijing 100080}\\
$^2${\small\it Department of Astronomy and Applied Physics, University of
                Science and Technology of China, Hefei, P. R. China
                \thanks{Mailing address}}\\
$^3${\small\it Department of Physics, Shanghai Jiaotong University,
                Shanghai, P. R. China}}
\maketitle

\begin{quotation}
    By the complex multimode Bogoliubov transformation, we obtain the
general forms of squeeze operators and squeezed states including squeezed
vacuum states, squeezed coherent states, squeezed Fock states and squeezed
coherent Fock states, for a general multimode boson system. We decompose the
squeezed operator into disentangling form in normal ordering to simplify the
expressions of the squeezed states. We also calculate the statistical
properties of the SS. Furthermore we prove that if it is non-degenerate, a
Bogoliubov transformation matrix can be decomposed into three basic matrix,
by which we can not only check the criterion of the minimum uncertainty
state (MUS) found by Milburn \cite{Mil}, but also prove that except the
special cases, any multimode squeezed state is MUS after the original
creation and annihilation operators rotate properly. We also discuss some
special cases of the three basic matrices. Finally we give an analytical
results of a two-mode system as an example.
\end{quotation}

\section{Introduction}

    Squeezed states(SS) are originally important concepts in quantum optics
\cite{Lou}. Because of the fascinating property that one of their two
quadratures has less uncertainty when compared with coherent states, SS have
several potential applications, such as in light communication,
amplification of signals, and gravitational radiation detection\cite{Wal}.
SS also apply to other quantum mechanical systems, such as polaritons and
polarons.

    Early researches focused only one and two-mode SS. Yuen \cite{Yue}
studied one-mode SS generally. Caves and Shumaker \cite{Cav} discussed
two-mode SS in a great detail. When we treat quantum mechanical systems
other than quantum optics, however, multimode SS inevitably come into sight.
Zhang, Feng and Gilmore \cite{Zha} and Lo and Sollie \cite{Los} put forward
multimode SS, but not in general form in fact. On the other hand, minimum
uncertainty states (MUS) is of basic interests in quantum physics. Stoler
\cite{Sto} found one-mode SS are MUS. And later Milburn \cite{Mil} proved
that multimode MUS are special multimode SS, which are simply direct
products of each one-mode squeezed state.

    Our present work is to construct the general multimode SS. For
convenience, we first give a short review of Fock states, coherent states
and coherent Fock states in Sec. II. In Sec. III we introduces complex
Bogoliubov transformation (BT) to diagonalize a positive definite bilinear
Hamiltonian. Based on the properties of the BT matrix, in Sec. IV we
construct concisely general multimode squeeze operator (SO), and decompose
it into three disentangling parts in normal ordering; then we compare our SO
with the usual one and two-mode forms. In Sec. V we present four kinds of SS,
squeezed vacuum states (SVS), squeezed coherent states (SCS), squeezed Fock
states (SFS), and squeezed coherent Fock states (SCFS), and their reduced
expressions; then we derive their covariance matrices, average quantum
numbers and their fluctuations; we calculate the wave functions of the SS at
the base of Fock state, coherent state respectively, and also in the
coordinate space. In Sec. VI we prove that if the BT matrix is
non-degenerate,  the BT matrix can be decomposed into three basic matrices,
each of which is also a BT matrix. Thus we can prove that SCS are MUS after
the original creation and annihilation operators rotate properly, In
addition, we analyze some interesting cases of the decomposed matrices,
including the case satisfying the criterion for multimode MUS given by
Milburn \cite{Mil}. To demonstrate our theory, in Sec. VII we give an
analytical results of a two-mode Hamiltonian as an example. Sec. VIII is the
conclusion and prospects of the further work. In Appendix A we give another
approach to obtain the SVS without the knowledge of the SO.

    In this paper index $T$ stands for transpose of a matrix, $*$ for
complex conjugate, and $\dagger$ for Hermite conjugate. $\sum_i$ ($\prod_i$)
means $\sum_i^N$ ($\prod_i^N$) unless stated specifically, and $D_f$ is the
short notation for an $N\times N$ diagonal matrix
Diag$\{f_1,f_2,\ldots,f_N\}$, where $f$ can be any symbol except $\alpha$,
since we use $D_\alpha$ to denote the Weyl displacement operator.

\section{Fock states, coherent states and coherent Fock states}

    For a system with Hamiltonian
$H_0=\sum_i\Omega_i(a_i^\dagger a_i+a_i a_i^\dagger)$, where $a_i$
($a_i^\dagger$) is the annihilation (creation) operator for the $i$-th
boson, the corresponding eigenstate is the Fock state
$|n\rangle=|n_1\rangle\otimes|0_2\rangle\otimes\cdots\otimes|n_N\rangle$.

    If adding a linear term to the above Hamiltonian, {\em i.e.},
$H=H_0+\sum_i\kappa_i a_i+\kappa^* a_i^\dagger$, the ground state is
a coherent state $|\alpha\rangle=D_\alpha|0\rangle$, where the displacement
operator $D_\alpha=e^{\sum_i(\alpha_i a_i^\dagger-\alpha_i^* a_i)}$,
and $\alpha_i=-{\kappa_i^*\over 2\Omega_i}$. The excited state is
$|n;\alpha\rangle=D_\alpha|n\rangle$, which we call the coherent Fock state.

    Fock state wave functions of the above three states are

\begin{equation}
\langle m|n\rangle=\prod_i\delta_{m_i, n_i},
\end{equation}

\begin{equation}
\langle m|\alpha\rangle=\prod_i e^{-{|\alpha_i|^2\over 2}}
   {\alpha_i^{m_i}\over\sqrt{m_i!}},
\end{equation}
and

\begin{equation}
\langle m|n;\alpha\rangle=\prod_i\left[{e^{-{|\alpha_i|^2\over 2}}\over
    \sqrt{m_i! n_i!}}\sum_{j=0}^{\min\{m_i, n_i\}}j! C_{m_i}^j C_{n_i}^j
    \alpha_i^{m_i-j}(-\alpha_i^*)^{n_i-j}\right].
\end{equation}

    Coherent state wave functions of the above three states are

\begin{equation}
\langle\beta|n\rangle=\prod_i e^{-{|\beta_i|^2\over 2}}
   {\beta_i^{*n_i}\over\sqrt{n_i!}},
\end{equation}

\begin{equation}
\langle\beta|\alpha\rangle=\prod_i e^{\beta_i^*\alpha_i
   -{1\over 2}(|\alpha_i|^2+|\beta_i|^2)}
\end{equation}
and

\begin{equation}
\langle\beta|n;\alpha\rangle
   =\prod_i{e^{\beta_i^*\alpha_i-{1\over 2}(|\alpha_i|^2+|\beta_i|^2)}
      (\beta_i^*-\alpha_i^*)^{n_i}\over\sqrt{n_i!}}.
\end{equation}

    We define two quadrature operators $X_i={a_i+a_i^\dagger\over 2}$, and
$Y_i={a_i-a_i^\dagger\over 2i}$, and ket
$|X\rangle=|X_1\rangle\otimes|X_2\rangle\otimes\cdots\otimes X_N\rangle$,
then the coordinate representation of the above three states are

\begin{equation}
\langle X|n\rangle=\prod_i({2\over\pi})^{1\over 4}
   {e^{-X_i^2}\over\sqrt{2^{n_i}n_i!}}H_{n_i}(\sqrt 2 X_i),
\end{equation}

\begin{equation}
\langle X|\alpha\rangle=\prod_i({2\over\pi})^{1\over 4}
    e^{-(X_i-\alpha_i)^2+{\alpha_i(\alpha_i-\alpha_i^*)\over 2}},
\end{equation}
and

\begin{equation}
\langle X|n;\alpha\rangle=\prod_i({2\over\pi})^{1\over 4}
   e^{-(X_i-\alpha_i)^2+{\alpha_i(\alpha_i-\alpha_i^*)\over 2}}
   {H_{n_i}(\sqrt 2 X_i-{\alpha_i+\alpha_i^*\over \sqrt 2})
     \over\sqrt{2^{n_i}n_i!}},
\end{equation}
where $H_{n_i}$ is the $n_i$-th Hermite polynomial.

    The covariance matrices of  the three states Cov$(X,Y)$ are

\begin{equation}
{1\over 4}\left[\begin{array}{cc}
  D_{2n+1}&iI_N\\-iI_N&D_{2n+1}\end{array}\right],~~~~~
{1\over 4}\left[\begin{array}{cc}I_N&iI_N\\-iI_N&I_N\end{array}\right],~~~~~
{1\over 4}\left[\begin{array}{cc}
  D_{2n+1}&iI_N\\-iI_N&D_{2n+1}\end{array}\right].
\end{equation}

    The average quantum numbers of  the three states $\langle N_i\rangle$
are

\begin{equation}
n_i,~~~~~|\alpha_i|^2,~~~~~n_i+|\alpha_i|^2.
\end{equation}

    The fluctuations of the quantum numbers of the three states
$\langle N_i^2\rangle$ are

\begin{equation}
0,~~~~~|\alpha_i|^2,~~~~~(2n_i+1)|\alpha_i|^2.
\end{equation}

    It is noticeable that  the coherent Fock state has the same covariant
matrix as the Fock state. This is because $D_\alpha$ only displaces the
operators by a constant, which has no contribution to the fluctuations.

\section{Bogoliubov Transformation}

    Bogoliubov transformation (BT), originally used in superconductivity
\cite{Bog}, can diagonalize a definite positive Hamiltonian of a multi-boson
system with its annihilation and creation operators in bilinear form.
Javannainen \cite{Jav} introduced a general {\em real} multimode BT for a
weakly interacting Bose-Einstein condensate confined to a spherically
symmetric harmonic potential. In order to construct {\em complex} SS, it is
necessary to extend Javannainen's theory to complex form.

    Suppose the Hamiltonian of the system is

\begin{eqnarray}
H&=&\sum_{i,j}a^\dagger_i\xi_{ij}a_j+a_i\xi_{ij}^* a^\dagger_j
    +a^\dagger_i\eta_{ij}a^\dagger_j+a_i\eta_{ij}^* a_j\nonumber\\
 &=&\left[\begin{array}{cc}a^{\dagger T}&a^T\end{array}\right]
    \left[\begin{array}{cc}\xi&\eta\\\eta^*&\xi^*\end{array}\right]
    \left[\begin{array}{c}a\\a^\dagger\end{array}\right]
\label{ha1}\end{eqnarray}
where both $\xi$ and $\eta$ are $N\times N$ matrices, and $a$ ($a^\dagger$)
is a column vector $(a_1,a_2,\ldots,a_N)^T$. Instead of using symmetric
Hamiltonian matrix in Ref. \cite{Jav}, we arrange the Hamiltonian matrix in
Hermitian form for convenience of the latter calculations.

    By the hermiticity of the Hamiltonian it is obvious that

\begin{equation}
\xi^\dagger=\xi,~~~~~\eta^T=\eta.
\end{equation}

    This Hamiltonian can be diagonalized when expressed by a new set of
boson operators $c_i$, $c^\dagger_i$, {\em i. e.},

\begin{equation}
H=\left[\begin{array}{cc}c^{\dagger T}&c^T\end{array}\right]
  \left[\begin{array}{cc}D_\Omega&0\\0&D_\Omega\end{array}\right]
  \left[\begin{array}{c}c\\c^\dagger\end{array}\right].
\label{ha2}\end{equation}

    The new boson operators relate to the old ones by a BT matrix $M$ as

\begin{equation}
   \left[\begin{array}{c}c\\c^\dagger\end{array}\right]
  =M\left[\begin{array}{c}a\\a^\dagger\end{array}\right]
  \equiv\left[\begin{array}{cc}u&v\\v^*&u^*\end{array}\right]
    \left[\begin{array}{c}a\\a^\dagger\end{array}\right],
\label{btm}\end{equation}
where both $u,~v$ are $N\times N$ matrices. By commutation relations
$[a_i,a_j^\dagger]=[c_i,c_j^\dagger]=\delta_{ij}$, $[a_i,a_j]=[c_i,c_j]=0$
and $[a_i^\dagger,a_j^\dagger]=[c_i^\dagger,c_j^\dagger]=0$, we obtain the
inverse of BT

\begin{equation}
   \left[\begin{array}{c}a\\a^\dagger\end{array}\right]
  =\left[\begin{array}{cc}u^\dagger&-v^T\\-v^\dagger&u^T\end{array}\right]
   \left[\begin{array}{c}c\\c^\dagger\end{array}\right]
  =KM^\dagger K\left[\begin{array}{c}c\\c^\dagger\end{array}\right],
\end{equation}
where $K=\left[\begin{array}{cc}I_N&0\\0&-I_N\end{array}\right]$, with $I_N$
the $N\times N$ identity matrix. Therefore $M^{-1}=KM^\dagger K$, {\em i.e.},

\begin{equation}
MKM^\dagger=M^\dagger KM=K,
\label{mkm}
\end{equation}
which in block form is

\begin{equation}
\left\{\begin{array}{l}uu^\dagger-vv^\dagger=I_N\\uv^T=vu^T
\end{array}\right.,\label{uud}\end{equation}
and

\begin{equation}
\left\{\begin{array}{l}u^\dagger u-v^T v^*=I_N\\u^\dagger v=v^T u^*
\end{array}\right..\label{udu}\end{equation}

    From the two representations of the Hamiltonian (\ref{ha1}) and
(\ref{ha2}), we have

\begin{equation}
    \left[\begin{array}{cc}\xi&\eta\\
      \eta^*&\xi^*\end{array}\right]
   =M^\dagger\left[\begin{array}{cc}D_\Omega&0\\
      0&D_\Omega\end{array}\right]M.
\label{mdm}\end{equation}

    Eq. (\ref{mdm}) multiplied on right by $KM^\dagger$, with the
consideration of the property of BT matrix $M$ (\ref{mkm}), yields

\begin{equation}
    \left[\begin{array}{cc}\xi&-\eta\\
      \eta^*&-\xi^*\end{array}\right] M^\dagger
   =M^\dagger\left[\begin{array}{cc}D_\Omega&0\\
      0&-D_\Omega\end{array}\right].
\label{eig}
\end{equation}

    Therefore the problem of diagonalizing the Hamiltonian is simplified to
be a familiar algebra problem of solving eigenvalues and eigenvectors of a
matrix. The eigenvectors are normalized by Eq. (\ref{mkm}).

    Define $\widetilde M=D_{ph}M$, with the phase matrix

\begin{equation}
   D_{ph}=\left[\begin{array}{cc}D_{e^{i\phi}}&0\\0&D_{e^{-i\phi}}
\end{array}\right].\end{equation}
Obviously, just like $M^\dagger$, $\widetilde M^\dagger$ also satisfies the
secular equation (\ref{eig}) and the normalization (\ref{mkm}). Thus for the
complex BT case, each eigenvector has an undetermined phase. While for the
real BT case as in Ref. \cite{Jav}, all the eigenvectors can be completely
determined up to the signs. Further study of such a property and its effects
may be very interesting.

    If the Hamiltonian (\ref{ha1}) is added by linear terms
$\sum_i\kappa_i a_i+\kappa_i^* a_i^\dagger$, {\em i. e.},

\begin{equation}
H=\left[\begin{array}{cc}a^{\dagger T}&a^T\end{array}\right]
    \left[\begin{array}{cc}\xi&\eta\\\eta^*&\xi^*\end{array}\right]
    \left[\begin{array}{c}a\\a^\dagger\end{array}\right]
  +\left[\begin{array}{cc}\kappa^T&\kappa^\dagger\end{array}\right]
    \left[\begin{array}{c}a\\a^\dagger\end{array}\right]
\end{equation}
 it can be diagonalized as

\begin{equation}
H=\left[\begin{array}{cc}c^{\dagger T}&c^T\end{array}\right]
    \left[\begin{array}{cc}D_\Omega&0\\0&D_\Omega\end{array}\right]
    \left[\begin{array}{c}c\\c^\dagger\end{array}\right]
  -{1\over 4}\left[\begin{array}{cc}\kappa^T&\kappa^\dagger\end{array}\right]
    \left[\begin{array}{cc}\xi&\eta\\\eta^*&\xi^*\end{array}\right]^{-1}
    \left[\begin{array}{c}\kappa^*\\\kappa\end{array}\right],
\end{equation}
 by a new transformation

\begin{equation}
   \left[\begin{array}{c}c\\c^\dagger\end{array}\right]
  =M\left(\left[\begin{array}{c}a\\a^\dagger\end{array}\right]
   +{1\over 2}
    \left[\begin{array}{cc}\xi&\eta\\\eta^*&\xi^*\end{array}\right]^{-1}
    \left[\begin{array}{c}\kappa^*\\\kappa\end{array}\right]\right).
\end{equation}

\section{Multimode Squeezed Operator}

\subsection{General case}

    The unitary SO $U$ should have the following property

\begin{equation}
   U\left[\begin{array}{c}a\\a^\dagger\end{array}\right]U^\dagger
  =M\left[\begin{array}{c}a\\a^\dagger\end{array}\right],
\label{uau}
\end{equation}
So for an arbitrary $2N\times 2N$ matrix $F$, we have

\begin{equation}
   U\left[\begin{array}{cc}a^{\dagger T}&a^T\end{array}\right]F
   \left[\begin{array}{c}a\\a^\dagger\end{array}\right]U^\dagger
  =\left[\begin{array}{cc}a^{\dagger T}&a^T\end{array}\right]M^\dagger F M
   \left[\begin{array}{c}a\\a^\dagger\end{array}\right].
\end{equation}
    Select $F$ satisfying

\begin{equation}
   M^\dagger FM=F,
\label{mfm}\end{equation}
Multiply Eq. (\ref{mfm}) on left by $MK$, with the use of (\ref{mkm}),
we get $KFM=MKF$, which means $KF$ and $M$ are commutative. Thus $KF$ is a
function of $M$, $F=Kf(M)$.

Consequently the SO can take the form

\begin{equation}
 U=e^{\left[\begin{array}{cc}a^{\dagger T}&a^T\end{array}\right]Kf(M)
   \left[\begin{array}{c}a\\a^\dagger\end{array}\right]}.
\end{equation}

    Suppose
$f=\left[\begin{array}{cc}f_{11}&f_{12}\\f_{21}&f_{22}\end{array}\right]$,
with all the four blocks being $N\times N$ matrices, it can be deduced that

\begin{equation}
   U\left[\begin{array}{c}a\\a^\dagger\end{array}\right]U^\dagger
  =e^{-\left[\begin{array}{cc}f_{11}-f_{22}^T&f_{12}+f_{12}^T\\
   f_{21}+f_{21}^T&f_{22}-f_{11}^T\end{array}\right]}
   \left[\begin{array}{c}a\\a^\dagger\end{array}\right].
\label{uaf}
\end{equation}
The form of $M$ in (\ref{btm}) suggests

\begin{equation}
   f_{12}=f_{21}^*,~~~~~f_{22}=f_{11}^*.
\end{equation}
And the unitarity of $U$ demands
$[K f(M)]^\dagger=-K f(M)$, namely

\begin{equation}
   \left[\begin{array}{cc}f_{11}^\dagger&-f_{21}^\dagger\\
   f_{12}^\dagger&-f_{22}^\dagger\end{array}\right]
  =\left[\begin{array}{cc}-f_{11}&-f_{12}\\f_{21}&f_{22}\end{array}\right].
\end{equation}
These two relations lead to

\begin{equation}
   f_{11}^T=-f_{22},~~~~~f_{12}^T=f_{12},~~~~~f_{21}^T=f_{21}.
\end{equation}
So (\ref{uaf}) is reduced to

\begin{equation}
   U\left[\begin{array}{c}a\\a^\dagger\end{array}\right]U^\dagger
  =e^{-2f(M)}\left[\begin{array}{c}a\\a^\dagger\end{array}\right].
\end{equation}
Compared with Eq. (\ref{uau}), we get

\begin{equation}
   f(M)=-{1\over 2}\ln M.
\end{equation}
Therefore the general complex SO of the $N$-mode boson system is

\begin{equation}
 U=e^{-{1\over 2}\left[\begin{array}{cc}a^{\dagger T}&a^T\end{array}\right]
   K\ln M\left[\begin{array}{c}a\\a^\dagger\end{array}\right]}.
\label{mso}
\end{equation}

    It is very interesting that the SO is in the same form in terms of the
new set of operators, {\em i. e.},

\begin{equation}
 U=e^{-{1\over 2}\left[\begin{array}{cc}c^{\dagger T}&c^T\end{array}\right]
   K\ln M\left[\begin{array}{c}c\\c^\dagger\end{array}\right]}.
\end{equation}

    This result can be proved directly by using the relation (\ref{mkm}),
but we can easily understand it by the definition of the SO (\ref{uau}),
since it is obvious that
$U\left[\begin{array}{c}c\\c^\dagger\end{array}\right]U^\dagger
=M\left[\begin{array}{c}c\\c^\dagger\end{array}\right]$ if we use the
relation (\ref{btm}), which is the same form as Eq. (\ref{uau}).

    Similarly, it can be proved that the product of two SO's

$$U_1=e^{-{1\over 2}\left[\begin{array}{cc}a^{\dagger T}&a^T\end{array}
    \right]K\ln (M_1)\left[\begin{array}{c}a\\a^\dagger\end{array}\right]}$$
and
$$U_2=e^{-{1\over 2}\left[\begin{array}{cc}a^{\dagger T}&a^T\end{array}
    \right]K\ln (M_2)\left[\begin{array}{c}a\\a^\dagger\end{array}\right]}$$
is a new SO as

\begin{equation}
U_1 U_2=
    e^{-{1\over 2}\left[\begin{array}{cc}a^{\dagger T}&a^T\end{array}\right]
    K\ln (M_2 M_1)\left[\begin{array}{c}a\\a^\dagger\end{array}\right]},
\label{u12}\end{equation}
minding the inverse order of the BT matrices $M_1$ and $M_2$.

    As best of our knowledge, our SO (\ref{mso}) is more general than all
previous forms. For example, Lo and Sollie \cite{Los} constructed a
multimode SO

$$S(\beta_{ij})=\exp\left[{1\over 2}\sum_{i\neq j}
    (\beta_{ij}a_i^\dagger a_j^\dagger-\beta_{ij}^*a_j a_j)\right].$$
Obviously, there are no $a_i a_j^\dagger$, $a_i^2$, and $a_i^{\dagger 2}$
terms in the exponent of this form. It results in

\begin{equation}
   U\left[\begin{array}{c}a\\a^\dagger\end{array}\right]U^\dagger
  =\left[\begin{array}{cc}\cosh|\beta|&{\sinh|\beta|\over|\beta|}\beta\\
   {\sinh|\beta|^T\over|\beta^T|}\beta^*&\cosh|\beta|^T\end{array}\right]
   \left[\begin{array}{c}a\\a^\dagger\end{array}\right],
\label{usp}\end{equation}
where $|\beta|$ is defined in Ref. \cite{Los}. Similar result was also
reported by Zhang, Feng and Gilmore \cite{Zha}, in which $a_i^2$ and
$a_i^{\dagger 2}$ terms are included.

    In Eq. (\ref{usp}) the diagonal blocks of the BT matrix are Hermitian,
and the two off-diagonal blocks are symmetric, which are not the necessary
properties of the four blocks defined in expression (\ref{btm}) for a
general BT matrix.

\subsection{One-mode and two-mode SO}

    It is interesting to compare our SO with the usual one-mode and two-mode
ones. For one-mode case, expression (\ref{mso}) is directly reduced to

\begin{equation}
   U^{(1)}=\exp\left({\ln\left({u+u^*\over 2}+\sqrt{({u+u^*\over 2})^2-1}~
   \right)\over 2\sqrt{({u+u^*\over 2})^2-1}}
   \left[\begin{array}{cc}a^{\dagger T}&a^T\end{array}\right]
   \left[\begin{array}{cc}{u^*-u\over 2}&-v\\
     v^*&{u^*-u\over 2}\end{array}\right]
   \left[\begin{array}{c}a\\a^\dagger\end{array}\right]\right).
\end{equation}

    Since there is one degree of freedom according to the discussion at the
end of Sec. III, we can select a real $u$, and thus get

\begin{equation}
   U^{(1)}=\exp({1\over 2}\zeta^*a^2-{1\over 2}\zeta a^{\dagger 2})
\end{equation}
where $\zeta={v\ln(u+|v|)\over|v|}$. Therefore the usual one-mode SO is
indeed a general one.

    We also prove that the usual two-mode SO

\begin{equation}
   U^{(2)}=\exp r(e^{-i\varphi}a_1 a_2-e^{i\varphi}a^\dagger_1 a^\dagger_2)
\end{equation}
can be deduced from expression (\ref{mso}) by selecting a {\em special} BT
with

$$u=\left[\begin{array}{cc}\cosh r&0\\0&\cosh r\end{array}\right],~~~~~
  v=\left[\begin{array}{cc}0&e^{i\varphi}\sinh r\\
    e^{i\varphi}\sinh r&0\end{array}\right].$$
Therefore the usual two-mode SO is {\em not} a general form.

\subsection{SO in disentangling form}
    Suppose the disentangling form of the SO in normal ordering to be

\begin{equation}
 U=C_0 e^{-a^{\dagger T}\rho a^\dagger}e^{-a^{\dagger T}\sigma a
   -a^T\sigma' a^\dagger}e^{a^T\tau a},
\label{dis}\end{equation}
where $C_0$ is a normalization coefficient, $\rho$, $\sigma$ and $\tau$ are
all $N\times N$ matrices. Furthermore, we can always arrange $\rho=\rho^T$,
$\sigma'=\sigma^T$, and $\tau=\tau^T$.

    It is straightforward that

 $$e^{a^T\tau a}\left[\begin{array}{c}a\\a^\dagger\end{array}\right]
   e^{-a^T\tau a}
  =\left[\begin{array}{cc}1&0\\2\tau&1\end{array}\right]
   \left[\begin{array}{c}a\\a^\dagger\end{array}\right],$$

 $$e^{-a^{\dagger T}\sigma a-a^T\sigma^T a^\dagger}
   \left[\begin{array}{c}a\\a^\dagger\end{array}\right]
   e^{a^{\dagger T}\sigma a+a^T\sigma^T a^\dagger}
  =\left[\begin{array}{cc}e^{2\sigma}&0\\0&e^{-2\sigma^T}\end{array}\right]
   \left[\begin{array}{c}a\\a^\dagger\end{array}\right],$$

 $$e^{-a^{\dagger T}\rho a^\dagger}
   \left[\begin{array}{c}a\\a^\dagger\end{array}\right]
   e^{a^{\dagger T}\rho a^\dagger}
  =\left[\begin{array}{cc}1&2\rho\\0&1\end{array}\right]
   \left[\begin{array}{c}a\\a^\dagger\end{array}\right].$$
So

\begin{eqnarray}
   U\left[\begin{array}{c}a\\a^\dagger\end{array}\right]U^\dagger
 &=&|C_0|^2\left[\begin{array}{cc}1&0\\2\tau&1\end{array}\right]
   \left[\begin{array}{cc}e^{2\sigma}&0\\0&e^{-2\sigma^T}\end{array}\right]
   \left[\begin{array}{cc}1&2\rho\\0&1\end{array}\right]
   \left[\begin{array}{c}a\\a^\dagger\end{array}\right]\nonumber\\
 &=&|C_0|^2\left[\begin{array}{cc}e^{2\sigma}&2e^{2\sigma}\rho\\
   2\tau e^{2\sigma}&4\tau e^{2\sigma}\rho+e^{-2\sigma^T}\end{array}\right]
   \left[\begin{array}{c}a\\a^\dagger\end{array}\right].
\end{eqnarray}
Comparing this result with (\ref{uau}), with the use of (\ref{uud}), we get

\begin{equation}
   |C_0|=1,~~~~~\rho={1\over 2}u^{-1}v,~~~~~\sigma={1\over 2}\ln u,
   ~~~~~\tau={1\over 2}v^*u^{-1}.
\label{rst}\end{equation}

    When there is no squeezing, $U=1$. On the other hand, $u=I_N$, $v=0$.
Thus $U=C_0$ according to Eq.'s (\ref{dis}) and (\ref{rst}). So $C_0=1$.

    This decomposition of SO is very useful in the calculation involving
SS, which will be shown in the next section.

\section{Four Kinds of Squeezed States}

\subsection{Expressions and reductions}

    SO is a key to SS, from which we can derive SVS, SCS, SFS and SCFS.
Physically, SVS (SFS) is the ground (excited) state of the Hamiltonian
(\ref{ha1}), and SCS (SCFS) is the ground (excited) state of that
Hamiltonian plus linear terms $\kappa_i^* a_i+\kappa_i a_i^\dagger$.
Respectively, the SVS, SCS, SFS and SCFS are

\begin{equation}
   |0\rangle_s=U|0\rangle,
\end{equation}

\begin{equation}
   |\alpha\rangle_s=e^{\sum_i\alpha_i c_i^\dagger-\alpha^*_i c_i}|0\rangle_s
  =U e^{\sum_i\alpha_i a^\dagger_i-\alpha^*_i a_i}|0\rangle=U|\alpha\rangle,
\end{equation}

\begin{equation}
   |n\rangle_s=\prod_i{(c_i^\dagger)^{n_i}\over\sqrt{n_i!}}|0\rangle_s
  =U\prod_i{(a^\dagger_i)^{n_i}\over\sqrt{n_i!}}|0\rangle
  =U|n_i\rangle,
\end{equation}
and

\begin{eqnarray}
   |n;\alpha\rangle_s
  &=&e^{\sum_i\alpha_i c_i^\dagger-\alpha^*_i c_i}
   \prod_i{(c_i^\dagger)^{n_i}\over\sqrt{n_i!}}|0\rangle_s\nonumber\\
  &=&U e^{\sum_i\alpha_i a^\dagger_i-\alpha^*_i a_i}
   \prod_i{(a^\dagger_i)^{n_i}\over\sqrt{n_i!}}|0\rangle
  =U|n;\alpha\rangle.
\end{eqnarray}
where

\begin{equation}
\left[\begin{array}{c}\alpha\\\alpha^*\end{array}\right]={1\over 2}
    \left[\begin{array}{cc}\xi&\eta\\\eta^*&\xi^*\end{array}\right]^{-1}
    \left[\begin{array}{c}\kappa^*\\\kappa\end{array}\right].
\end{equation}

    These formal expressions are symmetric for the annihilation and creation
operators, and are convenient such as for successive different SO's acting
on these states, according to the rule (\ref{u12}). But when calculating the
wave functions which we will do later, such forms are not so convenient. So
it is meaningful to use the disentangle formula (\ref{dis}) to simplify the
above four expressions, with only creation operators left.

    From (\ref{dis}) and (\ref{rst}), together with the identity
$e^{{\rm Tr}A}=|e^A|$, the SVS can be reduced as

\begin{eqnarray}
   |0\rangle_s
  &=&e^{-a^{\dagger T}\rho a^\dagger}e^{-a^{\dagger T}\sigma a
   -a^T\sigma^T a^\dagger}e^{a^T\tau a}|0\rangle\nonumber\\
  &=&|u|^{-{1\over 2}}e^{-a^{\dagger T}\rho a^\dagger}|0\rangle.
\label{ss1}\end{eqnarray}

    It is interesting that we can also get it in a totally different way
demonstrated in Appendix A.

    Since the energy of the SVS is Tr$D_\Omega$, and the average energy
of the original vacuum $\langle 0|H|0\rangle$is
Tr$\xi=$Tr$(uD_\Omega u^\dagger+v^*D_\Omega v^T)=$Tr
$[(u^\dagger u+v^T v^*)D_\Omega]=$Tr$[(1+2v^T v^*)D_\Omega]>
$Tr$D_\Omega$,
we emphasize an important property that the SVS are more stable than the
original vacuum states. In another word, the original vacuums do not exist
in a stationary system. This point is obvious physically because it is the
SVS other than the original vacuum states that are the ground states of the
Hamiltonian (\ref{ha1}).

    The SCS can be reduced as follows.

\begin{eqnarray}
   |\alpha\rangle_s
 &=&U e^{\alpha^T a^\dagger-\alpha^\dagger a}U^\dagger U|0\rangle\nonumber\\
 &=&e^{\left[\begin{array}{cc}-\alpha^\dagger&\alpha^T\end{array}\right]
   M\left[\begin{array}{c}a\\a^\dagger\end{array}\right]}
   |u|^{-{1\over 2}}e^{-a^{\dagger T}\rho a^\dagger}|0\rangle\nonumber\\
 &=&|u^*|^{-{1\over 2}}e^{-a^{\dagger T}\rho a^\dagger}
   e^{\left[\begin{array}{cc}-\alpha^\dagger&\alpha^T\end{array}\right]
   M\left[\begin{array}{cc}1&-2\rho\\0&1\end{array}\right]
   \left[\begin{array}{c}a\\a^\dagger\end{array}\right]}|0\rangle.
\nonumber\end{eqnarray}
Using (\ref{uud}) and (\ref{rst}), we find

 $$M\left[\begin{array}{cc}1&-2\rho\\0&1\end{array}\right]
  =\left[\begin{array}{cc}u&0\\v^*&u^{-T}\end{array}\right].$$
So

\begin{eqnarray}
   |\alpha\rangle_s
 &=&|u|^{-{1\over 2}}e^{-a^{\dagger T}\rho a^\dagger}
   e^{\left[\begin{array}{cc}-\alpha^\dagger&\alpha^T\end{array}\right]
   \left[\begin{array}{cc}u&0\\v^*&u^{-T}\end{array}\right]
   \left[\begin{array}{c}a^\dagger\\a\end{array}\right]}|0\rangle\nonumber\\
 &=&|u|^{-{1\over 2}}
   e^{-{1\over 2}\alpha^\dagger\alpha+\alpha^T\tau\alpha}
   e^{\alpha^T u^{-T}a^\dagger-a^{\dagger T}\rho a^\dagger}
   |0\rangle\nonumber\\
 &=&e^{-{1\over 2}\alpha^\dagger\alpha+\alpha^T\tau\alpha}
   e^{\alpha^T u^{-T}a^\dagger}|0\rangle_s.
\end{eqnarray}

    The reduction of the SFS is indirect and a little more complicated.
First,

\begin{eqnarray}
   |n\rangle_s
 &=&U\prod_i{(a^\dagger_i)^{n_i}\over\sqrt{n_i!}}U^\dagger U|0\rangle
   \nonumber\\
 &=&\prod_i{(\left[\begin{array}{cc}v^*&u^*\end{array}\right]
   \left[\begin{array}{c}a\\a^\dagger\end{array}\right])^{n_i}_i
   \over\sqrt{n_i!}}
   |u|^{-{1\over 2}}e^{-a^{\dagger T}\rho a^\dagger}|0\rangle\nonumber\\
 &=&|u|^{-{1\over 2}}e^{-a^{\dagger T}\rho a^\dagger}\prod_i
   {(v^* a+u^{-T}a^\dagger)^{n_i}_i\over\sqrt{n_i!}}|0\rangle.\nonumber
\end{eqnarray}

    Since $(v^* a+u^{-T}a^\dagger)_i$ and $(v^* a+u^{-T}a^\dagger)_j$
commute for different $i$ and $j$,

\begin{eqnarray}
e^{(v^*a+u^{-T}a^\dagger)^{n_i}_i}|0\rangle
  &=&{d^{n_i}\over dp^{n_i}}\left.
   e^{p^T(v^*a+u^{-T}a^\dagger)}|0\rangle\right|_{p_i=0}\nonumber\\
 &=&\left.{d^{n_i}\over dp^{n_i}}
   e^{p^T\tau p+p^T u^{-T}a^\dagger}|0\rangle\right|_{p_i=0},\nonumber
\end{eqnarray}
where $p=[p_1,p_2,\ldots,p_N]^T$. Therefore

\begin{eqnarray}
   |n\rangle_s&=&|u|^{-{1\over 2}}e^{-a^{\dagger T}\rho a^\dagger}
   \prod_i\left.{d^{n_i}\over dp^{n_i}}e^{p^T\tau p+p^T u^{-T}a^\dagger}
   |0\rangle\right|_{p_i=0}\nonumber\\
  &=&\prod_i{1\over\sqrt{n_i!}}\left.{d^{n_i}\over dp^{n_i}}
    e^{p^T\tau p+p^T u^{-T}a^\dagger}|0\rangle_s\right|_{p_i=0}.
\end{eqnarray}

    Based on the above skills, the SCFS  can be reduced to

\begin{eqnarray}
   |n;\alpha\rangle_s
 &=&|u|^{-{1\over 2}}e^{-a^{\dagger T}\rho a^\dagger}
    e^{\left[\begin{array}{cc}-\alpha^\dagger&\alpha^T\end{array}\right]
    \left[\begin{array}{cc}u&0\\v^*&u^{-T}\end{array}\right]
    \left[\begin{array}{c}a\\a^\dagger\end{array}\right]}\nonumber\\
  &&\mbox{}\hskip .5cm\prod_i{(v^*a+u^{-T}a^\dagger)^{n_i}_i\over\sqrt{n_i!}}
    |0\rangle\nonumber\\
 &=&|u|^{-{1\over 2}}e^{-a^{\dagger T}\rho a^\dagger}
    e^{\alpha^T u^{-T}a^\dagger-(\alpha^\dagger u-\alpha^T v^*)a}\nonumber\\
  &&\mbox{}\hskip .5cm\prod_i{1\over\sqrt{n_i!}}\left.{d^{n_i}\over dp^{n_i}}
   e^{p^T\tau p+p^T u^{-T}a^\dagger}|0\rangle\right|_{p_i=0}\nonumber\\
 &=&|u|^{-{1\over 2}}e^{-{1\over 2}\alpha^\dagger\alpha+\alpha^T\tau\alpha}
      e^{\alpha^T u^{-T}a^\dagger-a^{\dagger T}\rho a^\dagger}\nonumber\\
  &&\mbox{}\hskip .5cm\prod_i{1\over\sqrt{n_i!}}\left.{d^{n_i}\over dp^{n_i}}
   e^{p^T\tau p+p^T (u^{-T}a^\dagger+2\tau\alpha-\alpha^*)}
   |0\rangle\right|_{p_i=0}\nonumber\\
 &=&\prod_i{1\over\sqrt{n_i!}}\left.{d^{n_i}\over dp^{n_i}}
   e^{p^T\tau p+p^T (u^{-T}a^\dagger+2\tau\alpha-\alpha^*)}
   |\alpha\rangle_s\right|_{p_i=0}.
\label{ss4}\end{eqnarray}

\subsection{Statistical properties}

    The covariant matrice Cov$(X,Y)$ of the SVS or SCS is

\begin{equation}
{1\over 4}\left[\begin{array}{cc}(u-v)^\dagger(u-v)&i(u-v)^\dagger(u+v)\\
    -i(u+v)^\dagger(u-v)&(u+v)^\dagger(u+v)\end{array}\right],
\end{equation}

    Cov$(X,Y)$ of the SFS or SCFS is

\begin{equation}
{1\over 4}\left[\begin{array}{cc}
     (u-v)^\dagger D_{n+1}(u-v)+(u-v)^T D_n(u-v)^*
    &i(u-v)^\dagger D_{n+1}(u+v)-i(u-v)^T D_n(u+v)^*\\
    i(u+v)^T D_n(u-v)^*-i(u+v)^\dagger D_{n+1}(u-v)
    &(u+v)^\dagger D_{n+1}(u+v)+(u+v)^T D_n(u+v)^*
    \end{array}\right].
\end{equation}

    For the similar reason to the non-squeezing case discussed in Sec. II,
where the Fock state and the coherent Fock state have the same Cov$(X,Y)$,
here the SVS and SCS (the SFS and SCFS) have the same Cov$(X,Y)$.

    The average quantum numbers $\langle N_i\rangle$ of the four kinds of
state are

\begin{equation}
(v^\dagger v)_{ii},
\end{equation}

\begin{equation}
(v^\dagger v)_{ii}+(CC^\dagger)_{ii},
\end{equation}

\begin{equation}
(u^\dagger D_n u)_{ii}+(v^\dagger D_{n+1}v)_{ii}
\end{equation}
and

\begin{equation}
(u^\dagger D_n u)_{ii}+(v^\dagger D_{n+1}v)_{ii}+(CC^\dagger)_{ii},
\end{equation}
where $C=u^T\alpha^*-v^\dagger\alpha$.

    The fluctuations of quantum numbers $\Delta N_i^2$ of the four kinds of
states are

\begin{equation}
(u^\dagger v)_{ii}(v^\dagger u)_{ii}
    +(u^\dagger u)_{ii}(v^\dagger v)_{ii},
\end{equation}

\begin{eqnarray}
&(u^\dagger v)_{ii}(v^\dagger u)_{ii}
    +(u^\dagger u)_{ii}(v^\dagger v)_{ii}
    +(u^\dagger u+v^\dagger v)_{ii}(CC^\dagger)_{ii}&\nonumber\\
    &-(u^\dagger v)_{ii}(CC^T)_{ii}
    -(v^\dagger u)_{ii}(C^*C^\dagger)_{ii},&
\end{eqnarray}

\begin{eqnarray}
&(u^\dagger D_{2n+1}v)_{ii}(v^\dagger D_{2n+1}u)_{ii}
    +(u^\dagger D_{n+1}u+v^\dagger D_n v)_{ii}
     (u^\dagger D_n u+v^\dagger D_{n+1}v)_{ii}&\nonumber\\
    &-(A+B)^\dagger D_{n(n+1)}(A+B)-2A^\dagger D_{n(n+1)}B&
\end{eqnarray}
and

\begin{eqnarray}
&(u^\dagger D_{2n+1}v)_{ii}(v^\dagger D_{2n+1}u)_{ii}
    +(u^\dagger D_{n+1}u+v^\dagger D_n v)_{ii}
     (u^\dagger D_n u+v^\dagger D_{n+1}v)_{ii}&\nonumber\\
    &-(A+B)^\dagger D_{n(n+1)}(A+B)-2A^\dagger D_{n(n+1)}B
    +(u^\dagger D_{2n+1}u+v^\dagger D_{2n+1}v)_{ii}(CC^\dagger)_{ii}&
    \nonumber\\
    &-(u^\dagger D_{2n+1}v)_{ii}(CC^T)_{ii}
    -(v^\dagger D_{2n+1}u)_{ii}(C^*C^\dagger)_{ii},&
\end{eqnarray}
where $A_{ij}=|u_{ij}|^2$, $B_{ij}=|v_{ij}|^2$.

\subsection{Wave functions}

    Start from (\ref{ss1}-\ref{ss4}), we can derive the wave functions of
the four kinds of SS at the base of Fock state, coherent state and
coordinate. Some of the results can be compared with the single mode case in
Ref. \cite{Yu1}.

    At the Fock state base, the wave functions of the four kinds of states
are respectively

\begin{equation}
\langle m|0\rangle_s=|u|^{-{1\over 2}}\left(\prod_i{1\over\sqrt{m_j !}}
    {d^{m_j}\over dq_j^{m_j}}\right)e^{-q^T\rho q}|_{q_j=0},
\end{equation}

\begin{equation}
\langle m|\alpha\rangle_s=|u|^{-{1\over 2}}
    e^{-{1\over 2}\alpha^\dagger\alpha+\alpha^T\tau\alpha}
    \left(\prod_j{1\over\sqrt{m_j !}}{d^{m_j}\over dq_j^{m_j}}\right)
    e^{-q^T\rho q+q^T u^{-1}\alpha}|_{q_j=0},
\end{equation}

\begin{equation}
\langle m|n\rangle_s=|u|^{-{1\over 2}}
    \left(\prod_{i,j}{1\over\sqrt{n_i ! m_j !}}
    {d^{n_i+m_j}\over dp_i^{n_i}dq_j^{m_j}}\right)
    e^{p^T\tau p+q^T u^{-1}p-q^T\rho q}|_{p_i=q_j=0}
\end{equation}
and

\begin{eqnarray}
\langle m|n;\alpha\rangle_s&=&|u|^{-{1\over 2}}
    e^{-{1\over 2}\alpha^\dagger\alpha+\alpha^T\tau\alpha}
    \left(\prod_i{1\over\sqrt{n_i ! m_j !}}
        {d^{n_i+m_j}\over dp^{n_i}dq^{m_j}}\right)\nonumber\\
    &&e^{p^T\tau p+q^T u^{-1}p-q^T\rho q+
        p^T(2\tau\alpha-\alpha^*)+q^T u^{-1}\alpha}|_{p_i=q_j=0}.
\end{eqnarray}

    At the coherent state base, the wave functions of the four kinds of
states are respectively

\begin{equation}
\langle\beta|0\rangle_s=|u|^{-{1\over 2}}
    e^{-{1\over 2}\beta^\dagger\beta-\beta^\dagger\rho\beta^*},
\end{equation}

\begin{equation}
\langle\beta|\alpha\rangle_s=|u|^{-{1\over 2}}
    e^{\beta^\dagger u^{-1}\alpha
        -{1\over 2}(\alpha^\dagger\alpha+\beta^\dagger\beta
        -2\alpha^T\tau\alpha+2\beta^\dagger\rho\beta^*)},
\end{equation}

\begin{equation}
\langle\beta|n\rangle_s=|u|^{-{1\over 2}}
    e^{-{1\over 2}\beta^\dagger\beta-\beta^\dagger\rho\beta^*}
    \left(\prod_i{1\over\sqrt{n_i !}}{d^{n_i}\over dp_i^{n_i}}\right)
    e^{p^T\tau p+p^T u^{-T}\beta^*}|_{p_i=0}
\end{equation}
and

\begin{eqnarray}
\langle\beta|n;\alpha\rangle_s&=&|u|^{-{1\over 2}}
    e^{\beta^\dagger u^{-1}\alpha
        -{1\over 2}(\alpha^\dagger\alpha+\beta^\dagger\beta
        -2\alpha^T\tau\alpha+2\beta^\dagger\rho\beta^*)}\nonumber\\
    &&\left(\prod_i{1\over\sqrt{n_i !}}{d^{n_i}\over dp_i^{n_i}}\right)
    e^{p^T\tau p+p^T(u^{-T}\beta^*-\alpha^*+2\tau\alpha)}|_{p_i=0}.
\end{eqnarray}

    At the coordinate base, the wave functions of the four kinds of states
are respectively

\begin{equation}
\langle X|0\rangle_s=|u-v|^{-{1\over 2}}({2\over\pi})^{N\over 4}
    e^{-X^T(u-v)^{-1}(u+v)X},
\end{equation}

\begin{equation}
\langle X|\alpha\rangle_s=|u-v|^{-{1\over 2}}({2\over\pi})^{N\over 4}
    e^{-X^T(u-v)^{-1}(u+v)X+2X^T(u-v)^{-1}\alpha
        -{1\over 2}\alpha^\dagger\alpha
        -{1\over 2}\alpha^T(u-v)^*(u-v)^{-1}\alpha},
\end{equation}

\begin{eqnarray}
\langle X|n\rangle_s&=&|u-v|^{-{1\over 2}}e^{-X^T(u-v)^{-1}(u+v)X}
    \nonumber\\
    &&\left(\prod_i({2\over\pi})^{1\over 4}{1\over\sqrt{n_i !}}
        {d^{n_i}\over dp^{n_i}}\right)
    e^{-{1\over 2}p^T(u-v)^*(u-v)^{-1}p+2p^T(u-v)^*X}|_{p_i=0}
\end{eqnarray}
and

\begin{eqnarray}
\langle X|n;\alpha\rangle_s&=&|u-v|^{-{1\over 2}}
    e^{-X^T(u-v)^{-1}(u+v)X+2X^T(u-v)^{-1}\alpha
        -{1\over 2}\alpha^\dagger\alpha
        -{1\over 2}\alpha^T(u-v)^*(u-v)^{-1}\alpha}\nonumber\\
    &&\left(\prod_i({2\over\pi})^{1\over 4}{1\over\sqrt{n_i !}}
        {d^{n_i}\over dp^{n_i}}\right)
    e^{-{1\over 2}p^T(u-v)^*(u-v)^{-1}p
    +p^T[2(u-v)^*X-(u-v)^*(u-v)^{-1}\alpha-\alpha^*]}|_{p_i=0}.
\end{eqnarray}

\section{Decomposition of the BT matrix}

\subsection{Decomposition and MUS}

    We can constructively prove that a general BT matrix can be expressed by
multiplication of three basic matrices.

    From the first equation of (\ref{uud}), Hermitian matrices $uu^\dagger$
and $vv^\dagger$ can be diagonalized by a unitary matrix $S_1$
simultaneously. So we have $S_1^\dagger uu^\dagger S_1=D_{\cosh^2 r}$ and
$S_1^\dagger uu^\dagger S_1=D_{\sinh^2 r}$. The former identity indicates
that $T_1\equiv D_{1\over\cosh r}S_1^\dagger u$ is a unitary matrix, which
is to say, $u=S_1 D_{\cosh r}T_1$. Similarly $v=S_1 D_{\sinh r}T_2$.

    Substituting the expressions of $u$ and $v$ into the first equation of
(\ref{udu}), we can get $D_{\sinh^2 r}T_1 T_2^T=T_1 T_2^T D_{\sinh^2 r}$.
If $\sinh^2 r_i$ values differently for different $i$, then the unitary
matrix $T_1 T_2^T$ must be also diagonal, {\em i. e.},
$T_2=D_{e^{i\varphi}}T_1^*$. As a result,
$v=S_1 D_{e^{i\varphi}\sinh r}T_1^*$. Then we have

\begin{equation}
M=\left[\begin{array}{cc}S_1&0\\0&S_1^*\end{array}\right]
  \left[\begin{array}{cc}D_{\cosh r}&D_{e^{i\varphi}\sinh r}\\
   D_{e^{-i\varphi}\sinh r}&D_{\cosh r}\end{array}\right]
  \left[\begin{array}{cc}T_1&0\\0&T_1^*\end{array}\right]
\label{dec}\end{equation}

    But we can go a step further to move the factors $e^{i\varphi_i}$ of the
middle matrix of the r.h.s. of (\ref{dec}) to the other two matrices. Since

\begin{equation}
\left[\begin{array}{cc}D_{\cosh r}&D_{e^{i\varphi}\sinh r}\\
   D_{e^{-i\varphi}\sinh r}&D_{\cosh r}\end{array}\right]
 =\left[\begin{array}{cc}D_{e^{i\varphi/2}}&0\\
   0&D_{e^{-i\varphi/2}}\end{array}\right]
  \left[\begin{array}{cc}D_{\cosh r}&D_{\sinh r}\\
   D_{\sinh r}&D_{\cosh r}\end{array}\right]
  \left[\begin{array}{cc}D_{e^{-i\varphi/2}}&0\\
   0&D_{e^{i\varphi/2}}\end{array}\right],
\end{equation}

\begin{equation}
M=\left[\begin{array}{cc}S&0\\0&S^*\end{array}\right]
  \left[\begin{array}{cc}D_{\cosh r}&D_{\sinh r}\\
   D_{\sinh r}&D_{\cosh r}\end{array}\right]
  \left[\begin{array}{cc}T&0\\0&T^*\end{array}\right]
\equiv M_S M_D M_T,
\end{equation}
where $S=S_1 D_{e^{i\varphi/2}}$, $T=D_{e^{-i\varphi/2}}T_1$.

    It is easy to check that all of the three factor matrices $M_S$, $M_D$
and $M_T$ are BT matrices. Meanwhile, $M_S$ and $M_T$ are also unitary
and $M_D$ is Hermitian.

    Consequently the SO is decomposed according to the rule (\ref{u12}) as

\begin{equation}
    U=U_T U_D U_S,
\end{equation}
where $U_T=|T|^{-{1\over 2}}e^{-a^{\dagger T}\ln T a}$,
$U_D=e^{{1\over 2}(a^T D_r a-a^{\dagger T} D_r a^\dagger)}$, and
$U_S=|S|^{-{1\over 2}}e^{-a^{\dagger T}\ln S a}$.

    Literally, the BT matrix is well known as a symplectic matrix, which is
not the ample condition for such a decomposition. The other one is that the
two diagonal blocks ($u$ and $u^*$) and the two off-diagonal blocks ($v$ and
$v^*$) of the BT matrix are complex conjugated respectively.

    There is a problem about the squeezing property of the SVS and SCS.
Actually, it is not always squeezed with respect to the original quantum
$a_i$ and $a_i^\dagger$ according to Table II. But if we define a set of
operators

\begin{equation}
   \left[\begin{array}{c}b\\b^\dagger\end{array}\right]
  =M_T\left[\begin{array}{c}a\\a^\dagger\end{array}\right],
\end{equation}
then the SVS and the SCS are always squeezed with respect to the new
operators, because the covariant matrix with respect to the new operators is

\begin{equation}
   {\rm Cov}(X,Y)={1\over 4}
     \left[\begin{array}{cc}D_{e^{2r}}&i\\-i&D_{e^{-2r}}\end{array}\right].
\end{equation}

    Unlike $c_i$, which is a linear combination of both $a_i$'s and
$a_i^\dagger\mbox{}$'s, here $b_i$ is a linear combination of only $a_i$'s,
$b=Ta$. Since $T$ is unitary, it means a rotation in the operator space.

    Following Artoni and Birman \cite{Art}, we name the quasi-particles
corresponding to the set of $b_i$'s and $b_i^\dagger$'s as the mixed bosons,
and the squeezing as {\em intrinsic} squeezing.

    Therefore we conclude that both the SVS and SCS are MUS with respect to
the mixed bosons.

\subsection{Some special cases}

    In this subsection we discuss some special cases of BT matrix.

1. If $T=I_N$, then from Table II, for the SVS and SCS,

$$\langle\Delta X_i^2\rangle
    ={1\over 4}[(SD_{\cosh r}-SD_{\sinh r})^\dagger
        (SD_{\cosh r}-SD_{\sinh r})]_{ii}={1\over 4}D_{e^{-2r}},$$

$$\langle\Delta Y_i^2\rangle
    ={1\over 4}[(SD_{\cosh r}+SD_{\sinh r})^\dagger
        (SD_{\cosh r}+SD_{\sinh r})]_{ii}={1\over 4}D_{e^{2r}}.$$
So the SVS and SCS are MUS with respect to the original operators, which is
just the case discussed by Milburn \cite{Mil}.

2. If $T=D_{e^{i\varphi/2}}$, then

$$\langle\Delta X_i^2\rangle
    ={1\over 4}(\cosh 2r_i-\cos\varphi_i\sinh 2r_i),$$

$$\langle\Delta Y_i^2\rangle
    ={1\over 4}(\cosh 2r_i-\cos\varphi_i\sinh 2r_i),$$
which is the same as the result of the general one-mode case.

3. If $T=S^\dagger$, then $u=u^\dagger$, $v=v^T$, {\em i. e.} $M^\dagger=M$,
and we can get
$U=e^{{1\over 2}(a^TS^*D_r S^\dagger a-a^{\dagger T}SD_r S^T a^\dagger)}$.
It is noticeable that there are no $a_i a_j^\dagger$ terms, which is just
the case discussed by Zhang, Feng and Gilmore \cite{Zha} and Lo and Sollie
\cite{Los}.

\section{A Two-mode Example}

    In this section we calculate an example of a two-mode system.

Supposing a system with the Hamiltonian
\begin{eqnarray}
H&=&45(a_1 a_1^\dagger+a_1^\dagger a_1+a_2 a_2^\dagger+a_2^\dagger a_2)
     +36(a_1^2+a_1^{\dagger 2}+a_2^2+a_2^{\dagger 2})\nonumber\\
  &&+14i(a_1^\dagger a_2-a_1 a_2^\dagger)
    -32i(a_1 a_2-a_1^\dagger a_2^\dagger),
\nonumber\end{eqnarray}
from which we get
$$\xi=\left[\begin{array}{cc}45&7i\\-7i&45\end{array}\right],~~~~~
  \eta=\left[\begin{array}{cc}36&16i\\16i&36\end{array}\right].$$

According to our theory, the Hamiltonian is diagonalized as

$$H=30(c_1 c_1^\dagger+c_1^\dagger c1)+12(c_2 c_2^\dagger+c_2^\dagger c2),$$
and the BT matrix block

$$u=\left[\begin{array}{cc}{7\over 3\sqrt{10}}&{3i\over\sqrt{10}}\\
    {5i\over 3\sqrt{2}}&{1\over\sqrt{2}}\end{array}\right],~~~~~
v=\left[\begin{array}{cc}{2\over 3\sqrt{10}}&{2i\over\sqrt{10}}\\
    {2i\sqrt{2}\over 3}&0\end{array}\right].$$

    It is obtained that
$$\ln M=\left[\begin{array}{cccc}
     0&i\mu_1&\mu_2&i\mu_3\\i\mu_1&0&i\mu_3&\mu_4\\
     \mu_2&i\mu_3&0&i\mu_1\\i\mu_3&\mu_4&i\mu_1&0\end{array}\right],$$
where
$\mu_1=\gamma_1\sec\gamma_2\csc\gamma_3$,
$\mu_2=\gamma_1\tan\gamma_2+\ln r$,
$\mu_3=-\gamma_1\sec\gamma_2\cot\gamma_3$,
$\mu_4=\gamma_1\tan\gamma_2-\ln r$,
with $\gamma_1=\cos^{-1}\sqrt{15+7\sqrt{15}\over 60}$,
$\gamma_2=\cos^{-1}\sqrt{45-7\sqrt{15}\over 30}$,
$\gamma_3=2\tan^{-1}\sqrt{3\sqrt 5}$, and $r=\sqrt{3\over\sqrt 5}$.

So
$$U=e^{i\mu_1(a_1^\dagger a_2+a_1 a_2^\dagger)
      +i\mu_3(a_1 a_2+a_1^\dagger a_2^\dagger)
      -{\mu_2\over 2}(a_1^2+a_1^{\dagger 2})
      -{\mu_4\over 2}(a_2^2+a_2^{\dagger 2})}.$$

    The SO in normal ordering is
$$\sqrt{3\sqrt 5\over 11}
   e^{-{1\over 22}(7a_1^{\dagger 2}+5a_2^{\dagger 2}
    +6ia_1^\dagger a_2^\dagger)}
   e^{-A}
   e^{-{1\over 11}(2a_1^2+3a_2^2+2i\sqrt{5}a_1 a_2)},$$
where
$A=\left[\begin{array}{cc}a_1^\dagger&a_2^\dagger\end{array}\right]
     \left[\begin{array}{cc}\ln\sqrt{3\sqrt 5\over 11}
         +\mu(7-3\sqrt 5)&18\mu i\\
         10\sqrt 5\mu i&\ln\sqrt{3\sqrt 5\over 11}-\mu(7-3\sqrt 5)
     \end{array}\right]
     \left[\begin{array}{c}a_1\\a_2\end{array}\right]$, and
$\mu={\tan^{-1}{\sqrt{222\sqrt 5-94}\over 7+3\sqrt 5}
    \over\sqrt{222\sqrt 5-94}}$.

    Consequently, the SVS is

\begin{equation}
|0\rangle_s=\sqrt{3\sqrt 5\over 11}
   e^{-{1\over 22}(7a_1^{\dagger 2}+5a_2^{\dagger 2}
    +6ia_1^\dagger a_2^\dagger)}|0\rangle
\end{equation}

    The rotation matrix
$S=\left[\begin{array}{cc}
    {\sqrt 5-1\over 2\sqrt 3}&{\sqrt 5+1\over 2\sqrt 3}i\\
    {\sqrt 5+1\over 2\sqrt 3}i&{\sqrt 5-1\over 2\sqrt 3}\end{array}\right],$
$T=\left[\begin{array}{cc}
   \sqrt{{1\over 2}+{1\over\sqrt 5}}&-\sqrt{{1\over 2}-{1\over\sqrt 5}}i\\
   -\sqrt{{1\over 2}-{1\over\sqrt 5}}i&\sqrt{{1\over 2}+{1\over\sqrt 5}}
   \end{array}\right],$
and the diagonal block
$D_{\cosh r}={\rm Diag}\{\sqrt{{5\over 3}+{2\over 3\sqrt 5}},
      \sqrt{{5\over 3}-{2\over 3\sqrt 5}}\}$,
$D_{\sinh r}={\rm Diag}\{\sqrt{{2\over 3}+{2\over 3\sqrt 5}},
      \sqrt{{2\over 3}-{2\over 3\sqrt 5}}\}$.
Consequently,
$$U=e^{i\tan^{-1}(\sqrt 5-2)(a_1^\dagger a_2+a_1 a_2^\dagger)}
    e^{{ r_1\over 2}(a_1^2-a_1^{\dagger 2})
        +{r_2\over 2}(a_2^2-a_2^{\dagger 2})}
    e^{-i\tan^{-1}{3+\sqrt 5\over 2}(a_1^\dagger a_2+a_1 a_2^\dagger)},$$
where $r_{1,2}=\ln\left(\sqrt{{5\over 3}\pm{2\over 3\sqrt 5}}+
       \sqrt{{2\over 3}\pm{2\over 3\sqrt 5}}\right)$.

\section{Conclusion}

    The major achievements in this paper are threefold. First we constructed
the general SO for a multimode boson system by using the properties of the
complex BT matrix. The crux of the derivation is that the $2N\times 2N$
matrix $M$ should be considered instead of the isolated $N\times N$ blocks
$u$ and $v$. Second, we obtained four kinds of SS, which are ground or
excited states of a bilinear Hamiltonian or the Hamiltonian plus linear
terms, and simplified them by using the disentangling form of the SO in
normal ordering. Third, we proved that any $M$ is the multiplication of
three special BT matrices if $uu^\dagger$ has no degenerate eigenvalues,
which leads to an important conclusion that generally SVS or SCS are MUS
with respect to a set of mixed bosons.

    We will study several physical systems using our multimode SS theory in
the near future, which would also help us learn more properties of general
or special multimode system.

    We would like to thank Dr. Zidan Wang and Dr. Shaolong Wan for useful
discussions.

\appendix
\section{Construct the SVS by the coherent state}

    The annihilation and creation operators act on $\langle\alpha|$ yielding
respectively

\begin{equation}
   \langle\alpha|a_i
  =({\partial\over\partial\alpha_i^*}+{\alpha_i\over 2})\langle\alpha|,
~~~~~\langle\alpha|a^\dagger_i
  =\langle\alpha|\alpha_i^*=-2{\partial\over\partial\alpha_i}\langle\alpha|.
\label{bra}\end{equation}
Defining $f=\langle\alpha|0\rangle_s$, then

\begin{equation}
  0=\langle\alpha|c|0\rangle_s
   =\langle\alpha|ua+va^\dagger|0\rangle_s
   =[u({\partial\over\partial\alpha^*}+{\alpha\over 2})+v\alpha^*]f.
\end{equation}
The above equation multiplied on left by $u^{-1}$ yields

\begin{equation}
   [{\partial\over\partial\alpha^*}+{\alpha\over 2}
   +u^{-1}v\alpha^*]f=0.
\label{paf}
\end{equation}
The second equation of (\ref{uud}) indicates that $2\rho=u^{-1}v$ is a
symmetric matrix, satisfying

\begin{equation}
   {\partial\over\partial\alpha_i^*}(\sum_{jk}\alpha_j^*\rho_{jk}\alpha_k^*)
  =2\rho_{ij}\alpha_j^*.
\end{equation}
Thus from (\ref{paf}) we get

\begin{equation}
 f=f_0(\alpha)e^{-{1\over 2}\alpha^\dagger\alpha-\alpha^\dagger\rho\alpha^*}.
\end{equation}
(\ref{bra}) leads to
${\partial\over\partial\alpha_i}f=-{\alpha_i^*\over 2}f$, so $f_0$ is just a
constant irrelevant with $\alpha_i$.

    The overcompleteness relation of the coherent state is

\begin{equation}
   \prod_i\int{d\alpha_i d\alpha_i^*\over 2\pi i}
   |\alpha\rangle\langle\alpha|=1.
\label{ocr}
\end{equation}
    So

\begin{equation}
   |0\rangle_s=\prod_i\int{d\alpha_i d\alpha_i^*\over 2\pi i}
   |\alpha\rangle\langle\alpha|0\rangle_s
  =f_0\prod_i\int{d\alpha_i d\alpha_i^*\over 2\pi i}
   e^{-\alpha^\dagger\alpha-\alpha^\dagger\rho\alpha^*}e^{\alpha^T a^\dagger}
   |0\rangle.
\end{equation}
Divide $\alpha$ into real and imaginary part, {\it i.e.} let $\alpha=x+iy$,
then the above equation becomes

\begin{equation}
   |0\rangle_s=f_0\prod_i\int{dx_i dy_i\over\pi}
   e^{-\left[\begin{array}{cc}x^T&y^T\end{array}\right]
   \left[\begin{array}{cc}1+\rho&-i\rho\\-i\rho&1-\rho\end{array}\right]
   \left[\begin{array}{c}x\\y\end{array}\right]}
   e^{\left[\begin{array}{cc}x^T&y^T\end{array}\right]
   \left[\begin{array}{c}a^\dagger\\i a^\dagger\end{array}\right]}|0\rangle.
\label{axy}
\end{equation}
Using the Gaussian integral formula with a little change from Ref.\cite{Neg}

\begin{equation}
   \int{dx_1\ldots dx_{2n}\over\pi^n}
   e^{-\sum_{i,j=1}^{2n}x_i A_{ij}x_j+\sum_{i=1}^{2n}x_i J_i}
  =|A|^{-{1\over 2}}e^{{1\over 4}\sum_{i,j=1}^{2n}J_i A_{ij}^{-1}J_j},
\label{int}
\end{equation}
together with the two identities

 $$\left|\begin{array}{cc}1+\rho&-i\rho
   \\-i\rho&1-\rho\end{array}\right|=1,~~~~~~~~
   \left[\begin{array}{cc}1+\rho&-i\rho\\
   -i\rho&1-\rho\end{array}\right]^{-1}
  =\left[\begin{array}{cc}1-\rho&i\rho\\i\rho&1+\rho\end{array}\right],$$
(\ref{axy}) can be reduced to

\begin{equation}
   |0\rangle_s=f_0\left|\begin{array}{cc}1+\rho&-i\rho\\
   -i\rho&1-\rho\end{array}\right|^{-{1\over 2}}
   e^{{1\over 4}\left[\begin{array}{cc}
     a^{\dagger T}&i a^{\dagger T}\end{array}\right]
   \left[\begin{array}{cc}1+\rho&-i\rho\\
   -i\rho&1-\rho\end{array}\right]^{-1}
   \left[\begin{array}{c}a^\dagger\\i a^\dagger\end{array}\right]}|0\rangle
  =f_0 e^{-a^{\dagger T}\rho a^\dagger}|0\rangle,
\end{equation}

    $f_0$ can be determined from the normalization of $|0\rangle_s=1$.

    Inserting (\ref{ocr}), and using (\ref{bra}) together with
$\langle 0|\alpha\rangle=e^{-{1\over 2}\alpha^T\alpha^*}$, we get

\begin{eqnarray}
   _s\langle 0|0\rangle_s
 &=&|f_0|^2\langle 0|e^{-a^T\rho^* a}e^{-a^{\dagger T}\rho a^\dagger}
    |0\rangle\nonumber\\
 &=&|f_0|^2\prod_i\int{d\alpha_i d\alpha_i^*\over 2\pi i}
   \langle 0|e^{-a^T\rho^* a}|\alpha\rangle\langle\alpha|
   e^{-a^{\dagger T}\rho a^\dagger}|0\rangle\nonumber\\
 &=&|f_0|^2\prod_i\int{d\alpha_i d\alpha_i^*\over 2\pi i}
   e^{-\alpha^T\alpha^*-\alpha^T\rho^*\alpha-\alpha^\dagger\rho\alpha^*}
  =1.
\end{eqnarray}

    Set $\alpha_i=x_i+iy_i$, the above equation is transformed into

\begin{equation}
   1=|f_0|^2\prod_i\int\left({dx_i dy_i\over\pi}\right)
   \exp\left(-\left[\begin{array}{cc}x^T&y^T\end{array}\right]
   \left[\begin{array}{cc}1+\rho+\rho^*&i\rho^*-i\rho\\
      i\rho^*-i\rho&1-\rho-\rho^*\end{array}\right]
   \left[\begin{array}{c}x\\y\end{array}\right]\right).
\end{equation}
This integral can be calculated according to (\ref{int}) with $J=0$.

\begin{equation}
   1=|f_0|^2\left|\begin{array}{cc}1+\rho+\rho^*&i\rho^*-i\rho
   \\i\rho^*-i\rho&1-\rho-\rho^*\end{array}\right|^{-{1\over 2}}.
\end{equation}

    Define unitary matrix
$V={1\over\sqrt 2}\left[\begin{array}{cc}1&1\\-i&i\end{array}\right]$, then

\begin{equation}
 |f_0|^{-2}
  =\left|V^\dagger\left[\begin{array}{cc}1+\rho+\rho^*&i\rho^*-i\rho
   \\i\rho^*-i\rho&1-\rho-\rho^*\end{array}\right]V\right|^{-{1\over 2}}
  =\left|\begin{array}{cc}1&2\rho\\2\rho^*&1\end{array}\right|^{-{1\over 2}}
  =|1-4\rho^*\rho|^{-{1\over 2}}.
\end{equation}
From (\ref{rst}) and (\ref{uud}), we have
$1-4\rho^*\rho=(u^T u^*)^{-1}$, so

\begin{equation}
   f_0=e^{i\phi}|u|^{-{1\over 2}}.
\end{equation}

So we deduced the same SVS as in Sec. III, except for a free phase $\phi$.

\end{document}